\documentclass[conference]{IEEEtran}
\ifCLASSINFOpdf
\usepackage[pdftex]{graphicx}
\usepackage{subcaption}
\graphicspath{{images/}}

\DeclareGraphicsExtensions{.pdf,.jpeg,.png}
\else
\usepackage[dvips]{graphicx}
\graphicspath{eps}
\DeclareGraphicsExtensions{.eps}
\fi

\usepackage{ulem}

\makeatletter
\def\endthebibliography{%
	\def\@noitemerr{\@latex@warning{Empty `thebibliography' environment}}%
	\endlist
}
\makeatother
\usepackage{fixltx2e}
\usepackage{stfloats}
\usepackage{float}
\usepackage{ulem}
\usepackage{enumerate}
\usepackage{setspace}
\usepackage{balance}
\usepackage{caption}
\usepackage{xcolor}
\usepackage{graphicx}
\usepackage{algpseudocode}
\usepackage{color}
\usepackage{multicol}
\usepackage{multirow}
\usepackage[linesnumbered,ruled,vlined]{algorithm2e}
\usepackage[flushleft]{threeparttable}

\usepackage{setspace}
\usepackage{balance}
\usepackage[utf8]{inputenc}

\usepackage[scaled]{beramono}
\usepackage[T1]{fontenc}
\usepackage{amsmath}
\usepackage{amssymb}
\DeclareMathOperator*{\argmax}{arg\,max}

\usepackage{setspace}
\usepackage{cite}
\usepackage{amsmath}
\usepackage{amsthm}
\usepackage{listings}
\usepackage{textcomp}
\usepackage[utf8]{inputenc}
\usepackage{tabulary}
\usepackage{enumitem}
\usepackage[hyphens]{url}
\usepackage{booktabs}

\SetKwInput{KwInput}{Input}
\SetKwInput{KwOutput}{Output}
\DeclareUnicodeCharacter{00A0}{ }

\usepackage{romannum}

\usepackage{tikz}

\IEEEoverridecommandlockouts

\begin{document}
	\title{Deep-Edge: An Efficient Framework for Deep Learning Model Update on Heterogeneous Edge} 	
%	\title{DeepEdge: Efficient Management of Heterogeneous Edge Resources for Deep Learning Model Update} 	
	
	% author names and affiliations	
	\author{
	\IEEEauthorblockN{Anirban Bhattacharjee\thanks{* These authors contributed equally.
		Work performed primarily during doctoral studies at Vanderbilt
			University.}
		\IEEEauthorrefmark{1}\IEEEauthorrefmark{4},
		Ajay Dev Chhokra\IEEEauthorrefmark{1}\footnotemark\value{thanks}\IEEEauthorrefmark{2},
		Hongyang Sun\IEEEauthorrefmark{2}, \\  
		Shashank Shekhar\IEEEauthorrefmark{3}, Aniruddha Gokhale\IEEEauthorrefmark{2}, Gabor Karsai\IEEEauthorrefmark{2}, Abhishek Dubey\IEEEauthorrefmark{2}}
	\IEEEauthorblockA{\IEEEauthorrefmark{2}EECS Dept, Vanderbilt University, Nashville, TN, USA,\IEEEauthorrefmark{3}Siemens Corporate Technology, Princeton, NJ, USA and \\ \IEEEauthorrefmark{4}National Institute of Standards and Technology(NIST), Gaithersburg, MD, USA
	\\
		 Email: \IEEEauthorrefmark{4}anirban.bhattacharjee@nist.gov; \IEEEauthorrefmark{2}\{ajay.d.chhokra; hongyang.sun; \\
                 a.gokhale; gabor.karsai, abhishek.dubey\}@vanderbilt.edu
                 and \IEEEauthorrefmark{3}shashankshekhar@siemens.com
               }
	}

\let\footnote\thanks

% Shreyas Ramakrishna\IEEEauthorrefmark{2},
% shreyas.ramakrishna;
% Shashank Shekhar\IEEEauthorrefmark{3},

%	% As a general rule, do not put math, special symbols or citations
%	% in the abstract

% make the title area
\maketitle
\begin{abstract}

Deep Learning (DL) model-based AI services are increasingly offered in a variety of predictive analytics services such as computer vision, natural language processing, speech recognition. However, the quality of the DL models can degrade over time due to changes in the input data distribution, thereby requiring periodic model updates. Although cloud data-centers can meet the computational requirements of the resource-intensive and time-consuming model update task, transferring data from the edge devices to the cloud incurs a significant cost in terms of network bandwidth and are prone to data privacy issues. With the advent of GPU-enabled edge devices, the DL model update can be performed at the edge in a distributed manner using multiple connected edge devices. However, efficiently utilizing the edge resources for the model update is a hard problem due to the heterogeneity among the edge devices and the resource interference caused by the co-location of the DL model update task with latency-critical tasks running in the background. To overcome these challenges, we present Deep-Edge, a load- and interference-aware, fault-tolerant resource management framework for performing model update at the edge that uses distributed training. This paper makes the following contributions. First, it provides a unified framework for monitoring, profiling, and deploying the DL model update tasks on heterogeneous edge devices. Second, it presents a scheduler that reduces the total re-training time by appropriately selecting the edge devices and distributing data among them such that no latency-critical applications experience deadline violations. Finally, we present empirical results to validate the efficacy of the framework using a real-world DL model update case-study based on the Caltech dataset and an edge AI cluster testbed.

\end{abstract}

\renewcommand\IEEEkeywordsname{Keywords}
\begin{IEEEkeywords}
	Resource Management, Deep Learning, Edge Computing, Performance Optimization, Interference-aware, Distributed Training
\end{IEEEkeywords}
	
%------------------------------------------------------------------’
%--------------- ABSTRACT -------------------------------------
%\input{abstract}
\normalem
%------------------------------------------------------------------’
%--------------- INTRODUCTION -------------------------------------
\section{Introduction}
\label{sec:introduction}
% \subsection{Emerging Trends}
The past decade has seen substantial progress in \textit{Deep Learning} (DL), particularly \textit{Deep Neural Networks} (DNNs), leading to its widespread adoption in various domains, such as medicine~\cite{milletari2016v}, geology~\cite{huang2017scalable} and vehicular navigation~\cite{chen2015deepdriving}. DL models are trained using a large amount of data and have outperformed previous AI-based approaches. However, training a DL model is a resource-intensive and time-consuming task. Various distributed DL frameworks, such as Tensorflow~\cite{abadi2016tensorflow}, MXNET~\cite{chen2015mxnet} and Ray~\cite{moritz2018ray}, have been developed to reduce the training time by distributing the training workload among multiple machines (cluster) consisting of one or more Graphics Processing Units (GPUs) or Application Specific Integrated Circuits (ASICs), such as Tensor Processing Units (TPUs). 

\begin{figure}[htb]
    \centering
    \includegraphics[width = 0.75\linewidth]{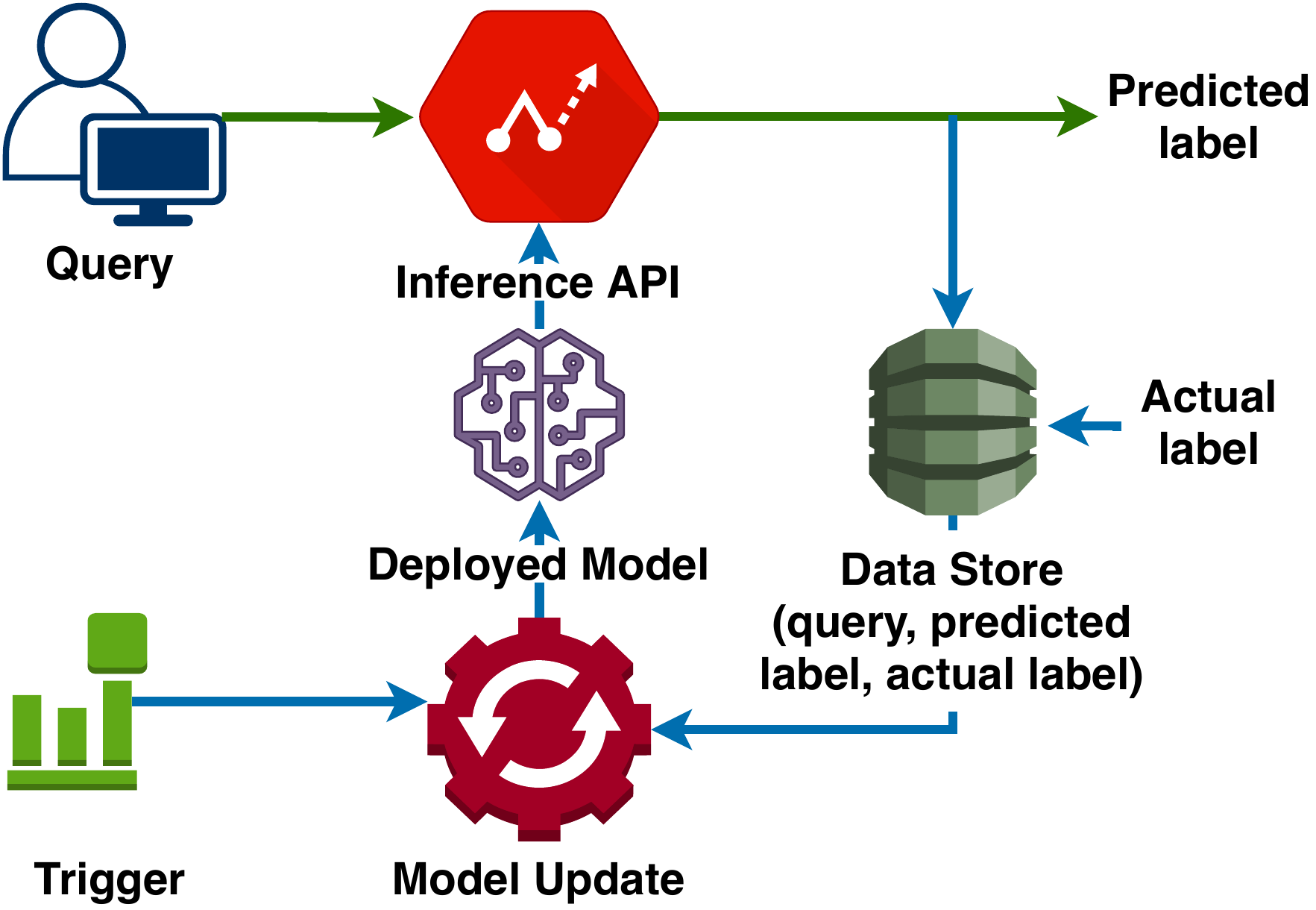}
    \caption{Machine learning model life cycle}
    \label{fig:lifecycle}
    \vskip -3mm
\end{figure}

However, an application composed of a deep learning component can experience degradation in accuracy over time due to changes in the input data distribution. 
% \anirban {
% Despite advances in DL technology, the predictive analytics-based application}
%that is composed of a deep learning component can experience a reduction in accuracy over time due to changes in the input data distribution. 
This phenomenon is referred to as \textit{Concept Drift}. To overcome model staleness and incorporate changes due to input data streams, \textit{continual learning} \cite{tian2018continuum} %\todo{[HS: I think we need a citation here -- Continuum]} 
has been used to periodically refine the static models by re-training the existing model using recent data. Figure~\ref{fig:lifecycle} shows the lifecycle of a machine learning model in production, where an inference API hosts the DL model. Recent data, along with the predicted and actual labels are stored in a data store that is fed to the model update process based on a user-defined trigger to replace the stale model with an updated one.

%Modified image a little bit

\subsection{Emerging Trends}
\label{subsec:emerging}
Traditionally, the model update occurs in the cloud, where the data collected from the edge of the network is transferred to the cloud data centers.  However, the availability of powerful and reliable GPU-accelerated edge AI products, such as NVIDIA's Jetson family (TX2, Nano, Xavier) and Google's edge TPU~\cite{hosseinabady2019heterogeneous} (Coral), has made continual learning viable using edge devices. In particular, the edge devices are suitable for performing the model update due to the following reasons:
\begin{itemize}
    \item The computational power of edge devices is sufficient for updating small to medium-sized DNNs (up to 150 million parameters), such as VGG, Resnet, Inception, Mobilenet.%, Densenet.
    \item The duration of the model update task is far less than the initial model training time, as fewer full data iterations (epochs) are required.  
    \item Performing model update at the edge avoids costly data transfers to the cloud.
    \item Using edge devices for the model update also handles data privacy concerns and reduces data security threats.
\end{itemize}

\subsection{Challenges}
\label{subsec:challenges}
There exist several resource managers (e.g., Borg~\cite{verma2015large}, Tetricsched~\cite{tumanov2016tetrisched}, Barista~\cite{bhattacharjee2019barista}) for different kinds of workloads in the cloud environment that perform static and dynamic scheduling. However, most of the approaches do not apply to edge clusters, which illustrate higher levels of heterogeneity among the edge devices. The heterogeneity can be the result of different physical characteristics of the devices, such as the number of processors, CUDA cores, memory, etc. or due to the workload associated with the devices. Another challenge is caused by the performance interference resulting from resource contention due to tasks running in the background. Running the resource-intensive model update task can cause the background latency-critical tasks to miss deadlines leading to Service-Level Objective (SLO) violations~\cite{bhattacharjee2019stratum}. Hence, the selection of an edge device to participate in the model update task should be contingent on the latency constraints of the background tasks. Resolving these challenges calls for a custom resource manager for DL model update workloads at the edge by considering the timing constraints of the background applications, the computational capabilities, and workloads of the individual edge devices along with the structure and characteristics of the DL jobs.

\subsection{Overview of Technical Contributions}
\label{subsec:overview}
In this paper, we propose \textit{Deep-Edge}, a custom resource management framework for DL model update jobs to minimize the model update time by distributing the re-training workloads among a set of heterogeneous edge devices while adhering to the timing and latency constraints of the background tasks. We focus on data-parallel distributed training based on the centralized parameter server architecture. Specifically, we make the following contributions:

\begin{itemize}
    \item We define unified monitoring, profiling, and deployment framework for model update tasks at the edge. 
    \item We build accurate performance and interference models for DL model update task and latency-critical background tasks by profiling them under various system metrics, such as CPU, GPU, Memory utilization, etc.
    \item We formulate an optimization problem that incorporates the edge node selection and workload distribution decisions to minimize the overall model update time.
    \item We present a polynomial-time heuristic solution based on the timing constraints, the performance, and interference models of the model update and background tasks.
    \item We show the efficacy of the framework by evaluating the accuracy of the proposed solution on a real-world DL model update task based on the Caltech dataset and an edge  AI  cluster testbed. 
    
\end{itemize}

\subsection{Paper Organization}
\label{subsec:organization}
The rest of the paper is organized as follows:
Section~\ref{sec:background} provides a brief introduction to DL and the challenges associated with updating a DL model using edge devices.
Section~\ref{sec:problem} describes the problem formulation.
Section~\ref{sec:approach} discusses the design and implementation of the \textit{Deep-Edge} framework.
Section~\ref{sec:evaluation} evaluates \textit{Deep-Edge} using a prototypical case study.
Section~\ref{sec:survey} presents a survey of existing solutions in the literature and compares them with \textit{Deep-Edge}.
Finally, Section~\ref{sec:conclusion} presents the concluding remarks and future directions.

%------------------------------------------------------------------’

%------------------------------------------------------------------’
%--------------- Motivation -------------------------------------
\section{Background and Motivation}
\label{sec:background}
Deep learning is the process of learning very complex functions by multiple transformations of the raw input to an abstract high-level representation~\cite{lecun2015deep}. Training a DL model such as a DNN is an iterative process that requires a large amount of data due to the number of parameters to be learned. The data is typically divided into \textit{shards}, which are further sliced up into \textit{batches}. The processing of a batch constitutes a training \textit{step}, which involves: 
1) inferring the output and calculating a loss function for each data sample in the batch (\textit{forward pass}); 
2) determining the gradients based on the loss function, i.e., changes to be made to the parameters of the DL model (\textit{backward pass}); and 
3) updating the parameters of the DL model for the given batch of samples. 
When all batches are processed, one \textit{epoch} is said to be completed.

Training a DL model is a resource-intensive and time-consuming task. Several machine learning frameworks, such as MXNET, TensorFlow, and Ray, support distributed training in which the task is divided among multiple workers. Primarily, there are two kinds of parallelism associated with distributed training: 
1) \textit{Model Parallelism} has all workers learn a part of the DL model parameters while working on the complete dataset; 
2) \textit{Data Parallelism} involves sharding the dataset among different workers such that each worker learns the complete DL model parameters while working on the part of the dataset. In this work, we focus on data-parallelism-based distributed training. 

There exists another classification in distributed training based on how the 
knowledge (parameters) learned by individual workers is shared across the group. Most DL frameworks implement either \textit{centralized} or \textit{decentralized} architecture for storing and sharing the updated parameters of a DL model. In the \textit{centralized} architecture, all workers compute forward and backward passes locally and send the gradients to a central entity, called the \textit{parameter server}, for updating the parameters based on an optimization algorithm such as \textit{Stochastic Gradient Descent (SGD)}. The parameters are then pulled back by the workers to continue the next training step. In the \textit{decentralized} architecture, no central entity exists, and the workers exchange among each other the locally learned gradients. The decentralized architecture is not suitable for the model update at the edge because it incurs higher transfer costs due to the need to broadcast the learned gradients to all the other workers.  

\begin{figure}[htb]
    \centering
    \includegraphics[width = 0.8\linewidth]{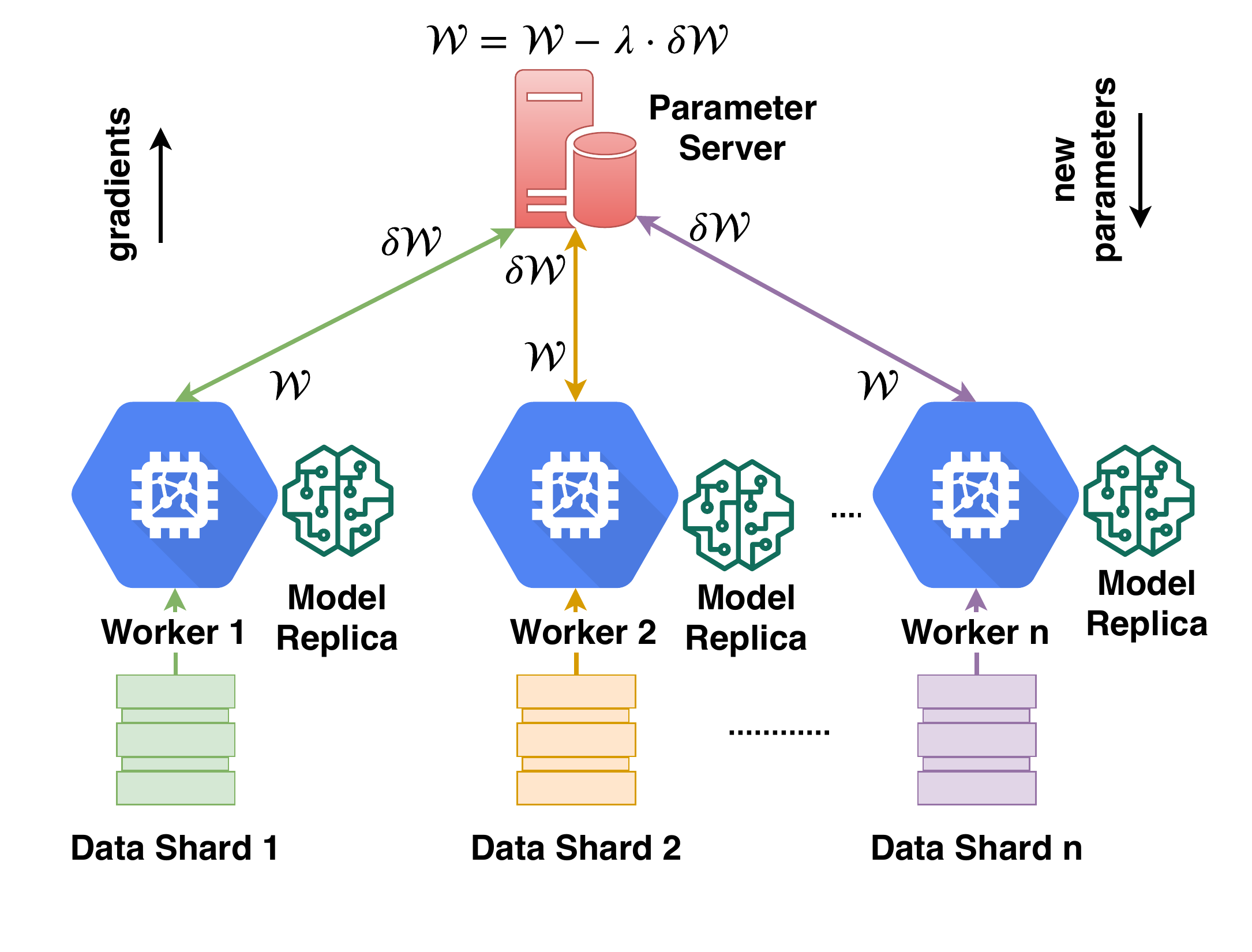}
    \caption{Parameter server architecture for distributed DL training}% with asynchronous training loops}
    \label{fig:ps}
    \vskip -4mm
\end{figure}

There are also two kinds of training loops associated with data-parallel, centralized distributed training: 
1) \textit{Synchronous training loop}, where each worker waits for the others to finish a training step before starting another step, i.e., the training progress is synchronized at each step; 
2) \textit{Asynchronous training loop}, where the training progress is not synchronized, and the parameter server updates the model parameters upon receiving the gradients from each worker. 
Using an asynchronous training loop is more favorable for the model update at the edge as it avoids the costly synchronization overhead. Figure~\ref{fig:ps} shows a centralized parameter server-based distributed training with $n$ asynchronous training loops.

The following subsections motivate the development of \textit{Deep-Edge} by describing the impact of heterogeneity and resource interference on the model update task and latency-critical tasks running in the background. 

\subsection{Impact of Heterogeneity on Model Update Time}

\begin{figure}[htb]
    \centering
    \includegraphics[width=\columnwidth]{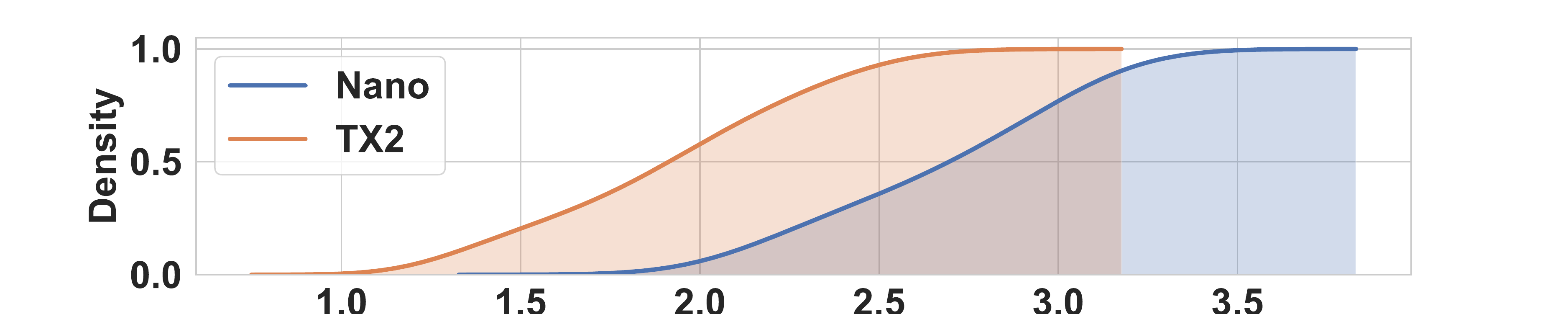}
    \caption{Variation of step time w.r.t device type}
    \label{fig:isolationcdf}
\end{figure}

\begin{figure}[htb]
    \centering
    \includegraphics[width=\columnwidth]{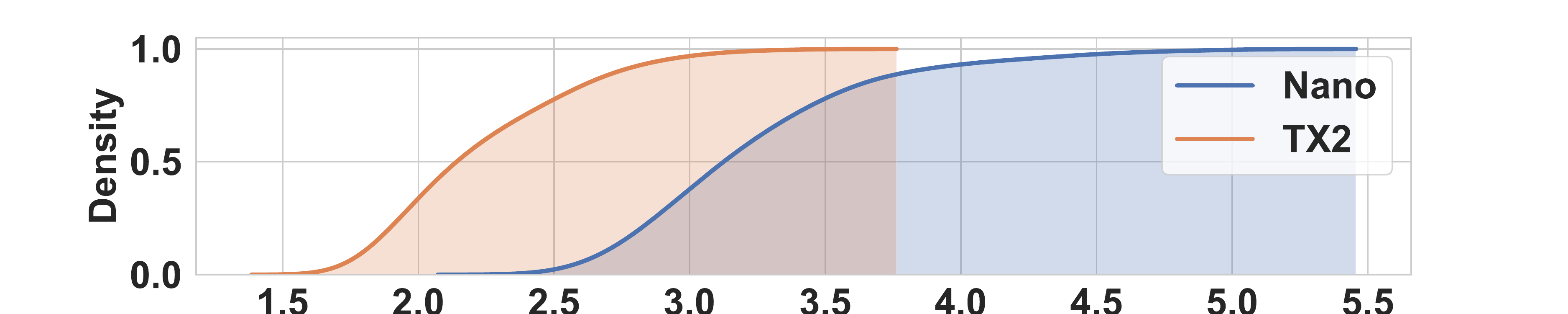}
    \caption{Increase in step time due to resource contention }
    \label{fig:gpucdf}
\end{figure}

Typically, an equal amount of data is distributed among the workers in multi-machine training. However, this approach can result in a longer time to complete because of the heterogeneity of the edge devices. For instance, we observed that TX2 is 30\% faster on an average in completing a training step than Nano when updating a state-of-the-art Inception model~\cite{szegedy2016rethinking} using the Caltech-256 Object category dataset~\cite{griffin2007caltech}. 
Figure~\ref{fig:isolationcdf} shows the cumulative distribution of time to complete one step, where the average step times for TX2 and Nano are 1.89 and 2.69 seconds, respectively. Thus, equal distribution can lead to underutilization of edge resources. Moreover, the performance of the model update task can be impacted by the state of the node (e.g., CPU, GPU, Memory utilization) as shown in Figure~\ref{fig:gpucdf}, where an initial GPU utilization of 88\% and 66\% for the two devices increases the average step time by almost 20\%. Hence, an intelligent data sharding policy is required by considering the actual states of the workers.

\subsection{Impact of Resource Interference on Background Tasks}
\begin{figure}[tbh]
    \centering
    \includegraphics[width = \linewidth]{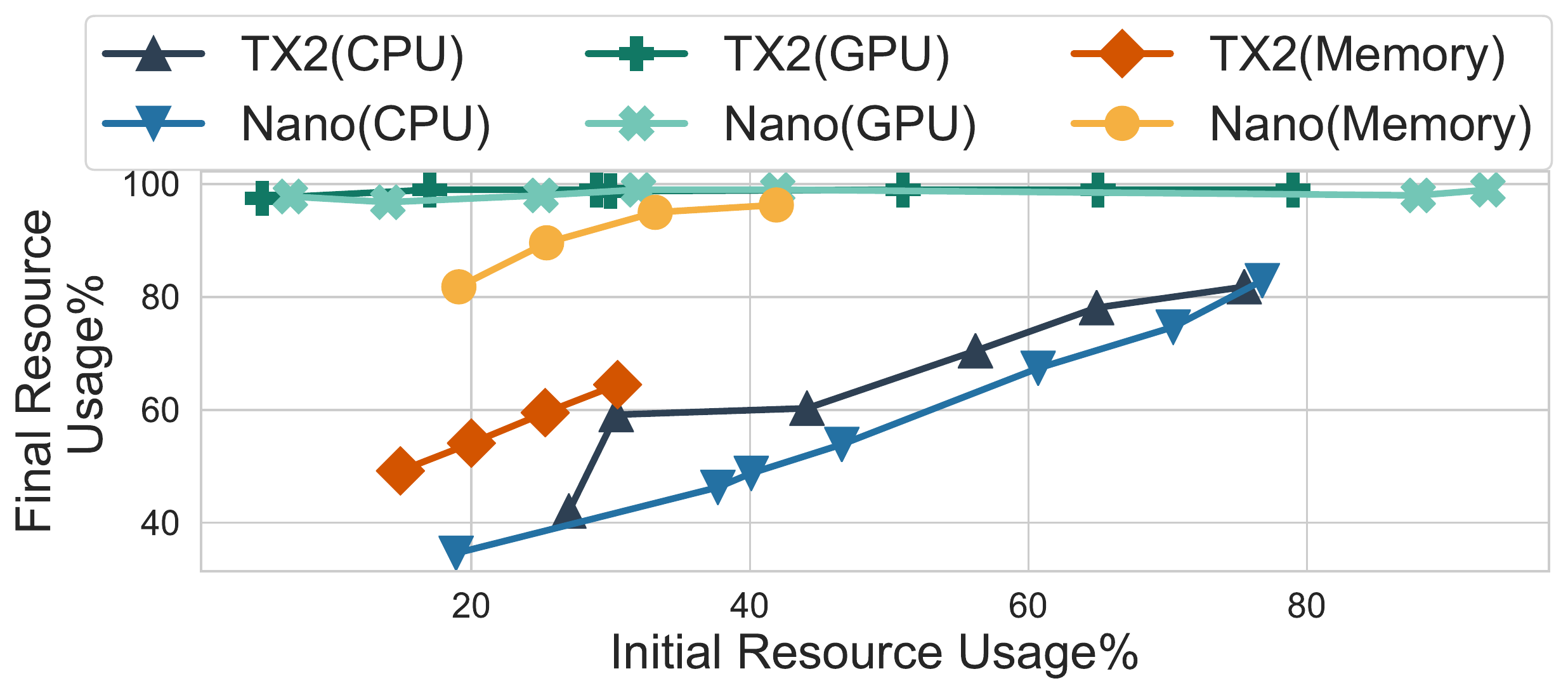}
    \caption{\scriptsize Initial and final resource usage along multiple resource dimensions.}
    \label{fig:Initital_Final_Features}
\end{figure}

The selection of a worker to participate in the model update task depends upon the degree of interference that can be tolerated by the worker's background tasks. Each curve in Figure~\ref{fig:Initital_Final_Features} shows the sensitivity of a resource (GPU, CPU, Memory) towards the DL model update task. The x-axis represents the initial resource utilization before running the DL update task, and the y-axis represents the final resource utilization. The updated system state (defined in terms of the system metrics) can lead to deadline violation of a latency-critical background task. Moreover, adding more workers can reduce the training throughput, as described in~\cite{peng2018optimus}. Hence, a resource scheduler for a DL task needs to find the optimal number of workers, along with the ideal data shards, without violating the SLO constraints of background tasks.

% \begin{figure}[t!]
%     \centering
%     \begin{subfigure}[t]{0.5\columnwidth}
%         \centering
%         \includegraphics[width = \columnwidth]{figures/pdfs/cdf_steptime_isolation.pdf}
%         \label{fig:imapcthetero}
%         \caption{Variation of step time w.r.t device type}
%     \end{subfigure}
%     ~ 
%     \begin{subfigure}[t]{0.5\columnwidth}
%         \centering
%         \includegraphics[width = \columnwidth]{figures/pdfs/cdf_steptime_gpu_5.pdf}
%         \caption{Performance degradation of model update task}
%     \end{subfigure}
%     ~
%     \begin{subfigure}[t]{\columnwidth}
%         \centering
%         % \includegraphics[width = \columnwidth]{images/dist_motivation1.pdf}
%         \caption{Performance of model updation task due to resource contention}
%     \end{subfigure}
%     \caption{Impact of Heterogeneity and Resource Interference}
% \end{figure}

%------------------------------------------------------------------’

%--------------- PROBLEMS -------------------------------------
\section{Problem Formulation}
\label{sec:problem}
In this paper, we consider distributed data-parallel training of DL models using a centralized parameter server architecture with asynchronous training loops. This section models the various costs involved in the DL model update process formulates an optimization problem to minimize the overall cost and states our assumptions.

\subsection{Cost Models}\label{sec:cost}
We consider a set $\mathcal{W} = \{w_1, w_2, \dots, w_N\}$ of $N$ heterogeneous edge nodes (or workers) that can perform distributed training, and a set $\mathcal{M}$ of data samples for training a DL model.
Let $\mathcal{D} = \{d_1, d_2, \dots, d_N\} $ denote the size of \emph{data shards} among all the workers, i.e., the number of data samples assigned to each worker, such that $\sum_{i=1}^{N} d_i = |\mathcal{M}|$.
%Let $\alpha_w \in [0, 1]$ denotes the proportion of data samples that are transferred from the server $s$ to an edge worker node $w$ for training, so node $w$ receives $\alpha_w \mathcal{M}$ data samples.

\subsubsection{Data transfer cost}
All data samples are assumed to be initially stored in a data store $ds \in \mathcal{DS}$, where $\mathcal{DS}$ is the set of data stores. The data samples need to be transferred to each edge worker for training.
Let $transfer^{ds}_i$ denote the cost of transferring a single data sample from data store $ds$ to edge node $w_i$. Thus, the total transfer cost for node $w_i$ is given by $Transfer^{ds}_i = d_i \cdot transfer^{ds}_i$.

\subsubsection{Initialization cost} After receiving the data samples, each edge node $w_i$ incurs a one-time initialization cost, denoted by $Initialize_i$, before the training begins. This cost is DL framework-dependent and consists mainly of data pre-processing (un-packing), loading the DL model, setting up the logical DL cluster, etc.

\subsubsection{Training cost} The data shards at each worker $w_i$ are further divided into \emph{batches} of size $b_i$, and let $\mathcal{B} = \{b_1, b_2, \dots, b_N\}$ denote the set of batch sizes for all workers.  The cost to process each batch includes the time to do forward propagation (for computing the loss function) and the time to do backward propagation (for computing the gradients). Let $forward_i$ denote the forward propagation time for one data sample on worker $w_i$, and let $backward_i$ denote the backward propagation time, which is typically incurred once per batch and is not related to the size of the batch. Given a batch size $b_i$, the per-sample compute time on worker $w_i$ is then given by $t_i^{compute} = forward_i + backward_i/b_i$.

After processing each batch, each worker $w_i$ pushes the gradients to a centralized parameter server $ps$ for update, and then pulls the updated parameters before continuing to train on the next batch. Let $push^{ps}_i$, $update^{ps}$ and $pull^{ps}_i$ denote the time to push, update and pull the parameters, respectively. Then, the update time is given by the sum of these three times, i.e., $t_i^{update} = push^{ps}_i + update^{ps} + pull^{ps}_i$. Since each worker $w_i$ has $B_i \approx d_i/b_i$ batches, the total time to process all the data samples on the worker, called an \emph{epoch}, is given by:
%\andy{I thought that due to heterogeneity, we were not dividing equally. HS: no we are not, and this equation incorporates heterogeneity in compute time $t^{compute}$$ and in data shard $d$}
\begin{align*}
epochTime_i &= B_i \Big(b_i \cdot t_i^{compute} + t_i^{update} \Big).
\end{align*}

Note that the per-sample compute time $t_i^{compute}$ to perform forward and backward propagation depends on the computing capability of the individual worker as well as the background tasks running on the worker. Further, the update time $t_i^{update}$ to perform push, update, and pull on each worker is also not fixed. It depends on the batch size, the total number of deployed workers, as well as the states of the workers and the parameter server.% that simultaneously perform the communication at any given time.
%We denote per sample compute time as $t_{compute, w}$, which is sum of $forward_w$, and $backward_w/b_w$.  The total time to synchronize the local parameters with the global parameters is denoted as $t_{update, w}$, which is the sum of $push^p_w$, $update^p$, $pull^p_w$ time.

\subsubsection{Total cost} The total cost includes the data transfer and initialization costs for all workers, followed by the asynchronous training and update costs from different workers. Note that all costs are expressed in terms of time. 

As the set of workers is assumed to be heterogeneous, some of them may not be deployed for training (e.g., due to high data transfer cost or low computational capability).
%Moreover, due to the asynchronous training, the number of epochs required for model convergence also depends upon the number of deployed workers and batch sizes.
Typically, having more workers will reduce the workload of each participating worker, thus decreasing the compute time (i.e., $t_i^{compute}$). However, it may also incur a larger update time (i.e., $t_i^{update}$) due to contentions caused by different workers trying to update the parameters at the same time. 

Let $\gamma_i \in \{0, 1\}$ denote a binary variable indicating if worker $w_i$ will be deployed for training or not, i.e., $\gamma_i = 1$ if $d_i > 0$ and $\gamma_i = 0$ if $d_i = 0$. Then, for all workers to complete a specified number of epoches, denoted by $numEpoch$, the total cost of distributed training can be expressed as:
\begin{align*}
Tota\_cost &= \max_i \{ \gamma_i \cdot \big( Transfer^s_i + Initialize_i \\ 
&~~~~~~~+ epochTime_i \cdot numEpoch \big) \}
% Total\_cost &= \max_i \{Transfer^s_i\} + \max_i \{Initialize_i \cdot \gamma_i\} \\
% &~~~~~~~+ \max_i \{epochTime_i\} \cdot numEpoch.
\end{align*}

% \todo{Hongyang: For the equation above, I think it would make more sense if there is only one max that includes the three costs for each server i, because as soon as a server receives the data, it can start initializing and training. Anirban and Ajay? }

%\todo{Number of epochs depends on batch size, data shard and number of workers.\\
%    HS: I've included both data shard $\alpha$ and batch size $b$ in the epoch expression above. Data shard $\alpha$ incorporates the number of workers implicitly since the ones not deployed will have $\alpha_w = 0$. }

\subsection{Optimization Problem}
%\andy{are we incorporating heterogeneity in this? HS: yes as mentioned before in the cost function.}

The goal of \emph{Deep-Edge} is to minimize the overall cost of distributed training on a set of heterogeneous edge nodes by choosing a data sharding scheme $\mathcal{D}$, the batch sizes $\mathcal{B}$, as well as the number of deployed workers while subject to some system and performance constraints. The following states the optimization problem:
\begin{align}
\text{minimize}\quad & Total\_cost \nonumber \\
\text{subject to}\quad & b_{\min} \le b_i \le b_{\max},~\forall i \label{eq:constraint1}\\
\quad & 0 \le d_i \le |\mathcal{M}|,~\forall i \label{eq:constraint2} \\
\quad & \sum\nolimits_{i} d_i = |\mathcal{M}| \label{eq:constraint3}  \\
\quad & pressure_i^a \leq \delta^a,~\forall a\in backApp_i, \forall i \label{eq:constraint4}
\end{align}

Constraint (\ref{eq:constraint1}) requires the batch size to be within the range of minimum and maximum system-specific batch size, which could be determined by the DL model or the device's memory constraint. Constraints (\ref{eq:constraint2}) and (\ref{eq:constraint3}) require that each worker receives a portion of the data samples, and altogether they cover the entire set of data samples. Finally, Constraint (\ref{eq:constraint4}) requires that, for each worker $w_i$, the pressure to its set of background applications, $backApp_i$, due to running the distributed training job on the same device, should be contained to be within an application-specific threshold $\delta_a$ for each application $a\in backApp_i$ in order not to violate the SLO of the application. The estimation of the pressure function to a background task, and the sensitivity function for the training job will be discussed further in Section \ref{sec:approach}.

As the objective function (i.e., $Total\_cost$) and the pressure constraint in the above optimization problem have complex, non-linear relationships with the decision variables (i.e., $\mathcal{D}, \mathcal{B}$), they cannot be expressed analytically. Thus the problem cannot be solved using standard solvers and/or analytical techniques. Therefore, we will design efficient heuristic solutions for the problem, which will be described in Section~\ref{sec:approach}.

\subsection{Assumptions}
% Because of the heterogeneity of the edge cluster and the background loads of the edge nodes, each worker may have a different per-sample compute time. Further, the update time of a worker may also vary if we change the total number of deployed workers. Therefore, we will build a performance model (Section \ref{sec:perf_model}) to estimate both the per-sample compute time and the update time for each worker in order to design an effective data sharding strategy (Section \ref{sec:sharding}).

We assume that the user specifies the maximum number of epochs (i.e., $numEpoch$) required for the training of the DL model, which is independent of the configuration of the workers.  The user also provides the model trigger condition, and the model is only updated whenever the trigger condition is received. As the DL model is usually updated on a large amount of data, we further assume that the one-time transfer cost (i.e., $Transfer^s_i$) and initialization cost (i.e., $Initialize_i$) are negligible compared to the total training cost. Finally, we assume that the location of the parameter server is given, and we do not need to select a node as a parameter server based on some specific criteria. 

%------------------------------------------------------------------’

%--------------- approach -------------------------------------
\section{Design and Implementation of Deep-Edge}
\label{sec:approach}
This section presents the design and implementation details of \textit{Deep-Edge} by describing the architecture model, various components of the framework, and its modes of operation.

\subsection{Architecture Model}
\label{sec:architecture}
\begin{figure}[htb]
    \centering
    \includegraphics[width = 0.8\linewidth]{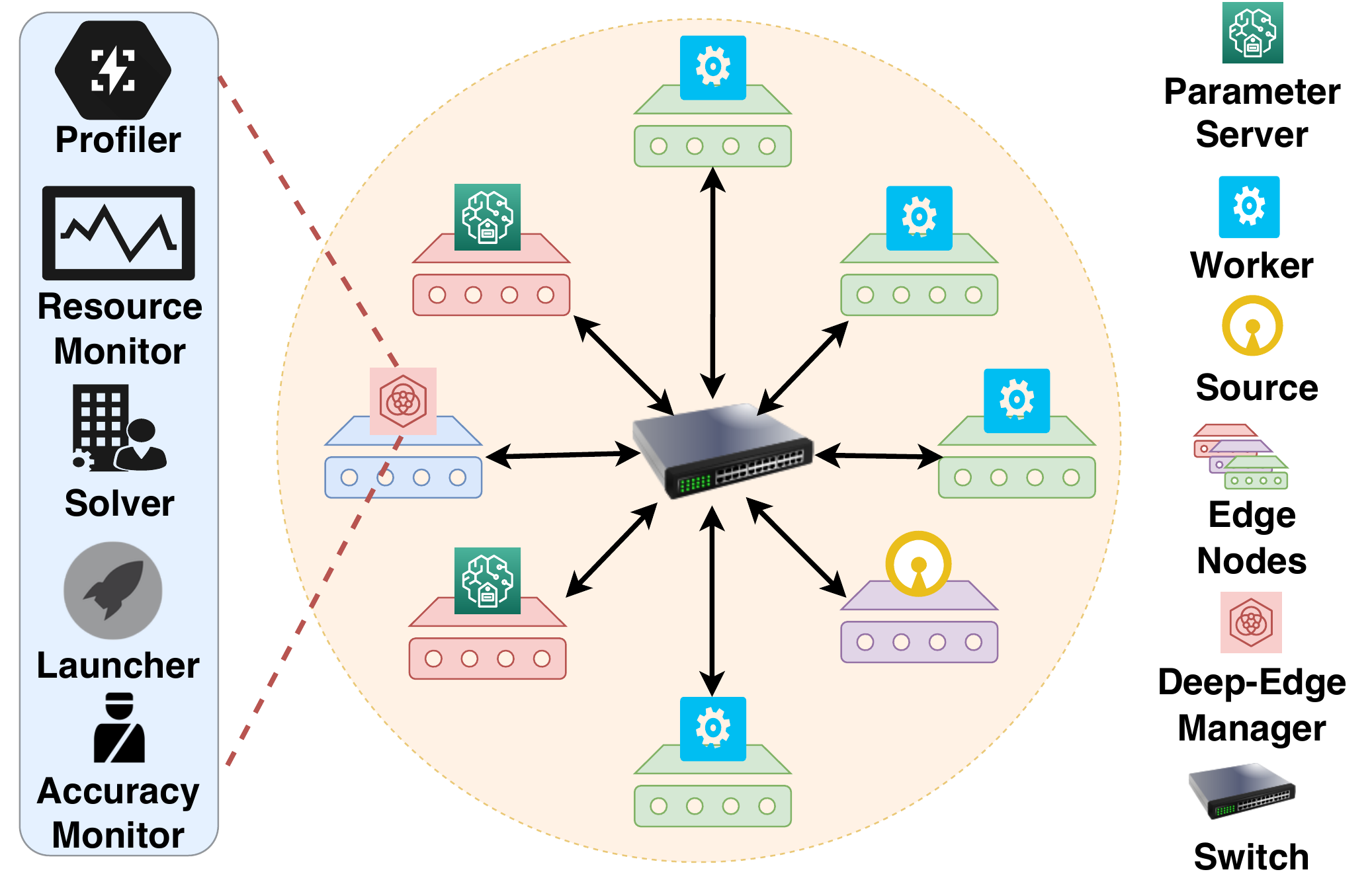}
    \caption{Deep-Edge architecture }
    \label{fig:arch}
    \vskip -5mm
\end{figure}

The architecture model is shown in Figure~\ref{fig:arch} consists of $K$ edge nodes, out of which $N$ are workers, $O$ are data stores, and $P$ are parameter servers, and they are represented by the disjoint sets $\mathcal{W}$, $\mathcal{DS}$ and $\mathcal{PS}$, respectively. These edge nodes form a local area network and are connected via a layer 2 switch. As mentioned in the previous section, the nodes in $\mathcal{W} \cup \mathcal{PS}$ form the DL model update cluster, where the worker nodes perform the actual re-training and the parameter server nodes act as a central repository for model parameters. The nodes in $\mathcal{DS}$ store data samples for the model update task. 

One of the nodes in $\mathcal{DS}$ also acts as a \textit{Deep-Edge Manager} (DEM), which implements the \textit{Deep-Edge} framework. DEM consists of a collection of components that enables profiling, resource scheduling and runtime monitoring of the DL cluster. The components are hosted as REST endpoints as \texttt{http://<ip>:<port>/<endpt\_name>?<endpt\_args>}, where \texttt{ip} is the IP address of the host, \texttt{port} is the port associated with the DEM, and \texttt{endpt\_name}, \texttt{endpt\_args} are the name and input arguments of the endpoint. The following subsections describe the different components and modes of operation associated with the DEM.

\subsection{Components of Deep-Edge Manager}
The DEM consists of five components, which together provide a unified solution for profiling, scheduling, and monitoring the DL model update tasks.
\subsubsection{\textbf{Profiler}} \textit{Deep-Edge} uses a data-driven approach to estimate various performance and interference models (i.e., $t^{compute}$, $t^{update}$  and $pressure$ experienced by the background applications). This component allows both latency-critical and model update tasks to be profiled against stress points along the dimensions of CPU, GPU, and Memory utilization. The Profiler accepts different system metrics and stress points as the input arguments. \textit{Deep-Edge} uses CPU, Memory, Disk I/O stressors from the well-known library Stress-ng~\cite{king2017stress}, and the GPU load stressing application is based on the NVIDIA Cuda-10 library. 
\subsubsection{\textbf{Solver}} This component implements the scheduling strategy to identify the candidate workers and their respective data shards. The Solver takes the number of data samples, the current state of the edge cluster along with the performance and interference models as inputs, and outputs a data sharding scheme. The scheduling strategy is explained in more detail in Section~\ref{subsec:scheduling}
\subsubsection{\textbf{Resource Monitor}} The Resource Monitor maintains a map of active workers in the edge cluster and periodically monitors the ongoing model update tasks. This component is responsible for re-triggering the model update task in response to worker node failure. 
\subsubsection{\textbf{Launcher}} This component is responsible for launching applications (DL \& stressing) on the worker and parameter server nodes. It accepts three different kinds of arguments: a) Stressor arguments; b) DL arguments; and c) Logging arguments. The Stressor arguments include parameters for the different stressing applications. The DL arguments include machine learning specific arguments, such as batch size, number of epochs, optimizer, etc. The Logging arguments include the file path for creating log files. 
\subsubsection{\textbf{Accuracy Monitor}} \textit{Deep-Edge} allows dynamic stopping of the model update task by tracking the accuracy of the validation set. The update task is launched based on the initial estimate of the number of epochs provided by the user, and the estimate of the number of epochs is refined in an online fashion using logistic regression, similar to Optimus~\cite{peng2018optimus}. 

\begin{figure}[!tbh]
    \centering
    \includegraphics[width = 0.9\linewidth]{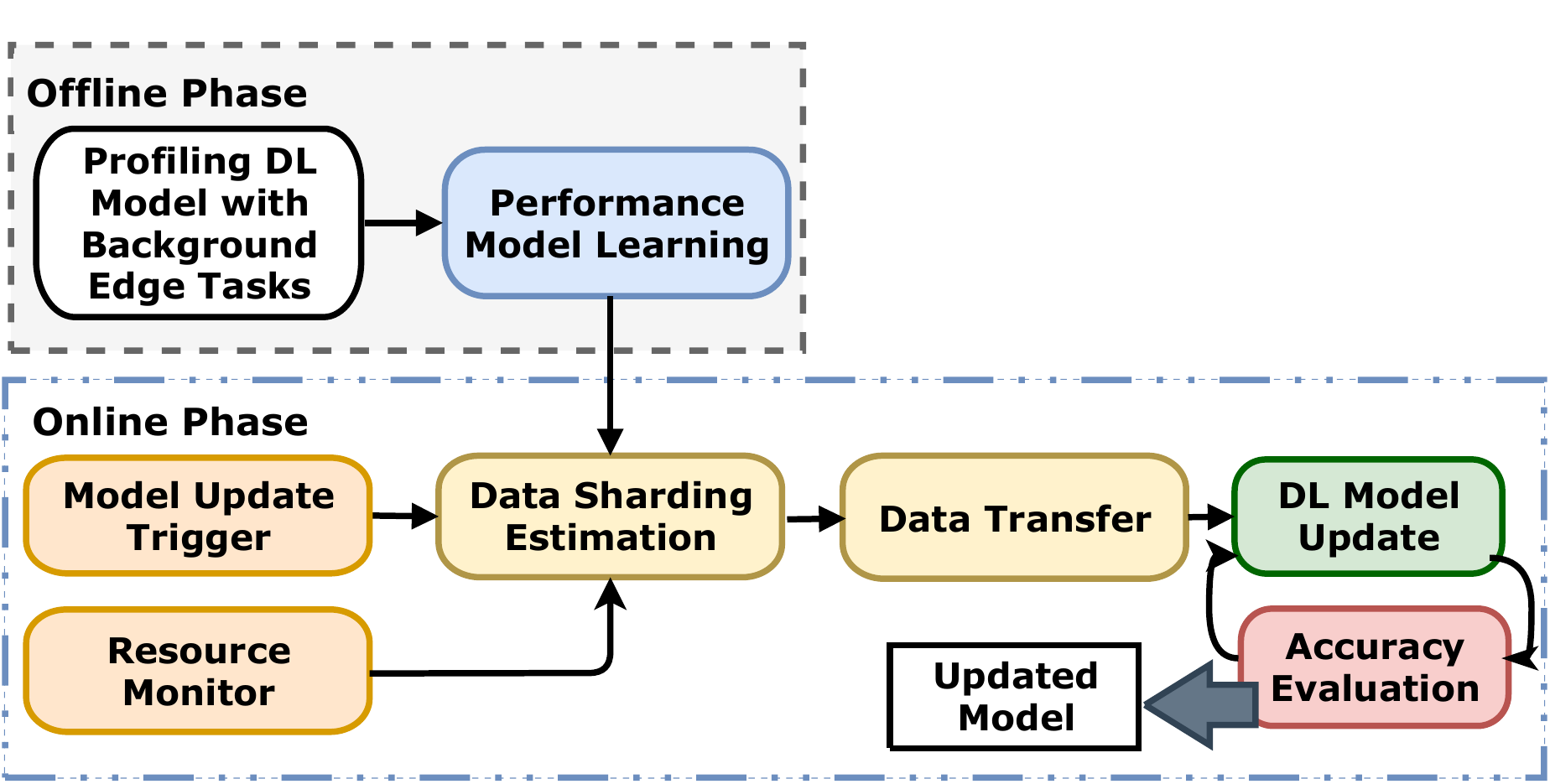}
    \caption{Modes of operation}
    \label{fig:mode}
\end{figure}

\subsection{Modes of Operation}
The operation of DEM can be categorized into two modes, \textit{offline} and \textit{online}, as illustrated in Figure~\ref{fig:mode}. In the \textit{offline} or \textit{design} mode, we build machine learning models to predict the execution times of the DL re-training task and latency-critical background tasks. In the \textit{online} mode, appropriate workers, along with batch size distribution and data shards, are calculated based on the developed performance models such that re-training time is minimized while adhering to the SLO constraints imposed by the background applications. The \textit{online} mode also handles worker failures by trying to restart the DL re-training task. The details of these features are described in the following. 

\subsubsection{\textbf{Performance and Interference Modeling}} 
\textit{Deep-Edge} uses a data-driven approach for modeling the performance of DL re-training tasks on each node as well as the interference experienced by latency-critical background tasks. The performance of the DL re-training task is measured by the time to complete one epoch, $epochTime_i$, which is the function of per-sample compute time, $t^{compute}_i$, and update time, $t^{update}_i$, as defined in the optimization problem. Here, $t^{compute}_i$ depends upon the node type, state and batch size. We describe node state as a vector of system metrics containing CPU, GPU and Memory utilizations. However, $t^{update}_i$ depends not only on the node state but also on the complete batch distribution, state of the parameter server and number of workers. We define two functions, \texttt{EstComputeTime} and \texttt{EstUpdateTime}, to model the relation between the state of the DL cluster and $t^{compute}_i$ and $t^{update}_i$ as shown in Equations~(\ref{eq:model1}) and (\ref{eq:model2}) below, where $\mathcal{X}_{i}$ and $b_i$ represent the state and batch size associated with worker $w_i \in \mathcal{W}$, $\mathcal{B}$ is the batch size distribution and $\mathcal{X}_{ps}$ is the state of the parameter server. 
\begin{align}
& t^{compute}_{i} = \texttt{EstComputeTime}(\mathcal{X}_{i}, b_i) \label{eq:model1} \\
& t^{update}_{i} = \texttt{EstUpdateTime}(\mathcal{X}_{i}, \mathcal{B}, \mathcal{X}_{ps}, |\mathcal{W}|) \label{eq:model2}
\end{align}

In order to create an interference profile of the DL re-training task, we model interference as performance degradation experienced by the background applications. We use a two-step approach to quantify performance degradation, i.e., increase in execution time. In the first step, we model the effects of running the DL re-training task on a node whose state is described in Equation~(\ref{eq:model3}). The function, \texttt{EstState}, gives the relation between the initial state of the node, $\mathcal{X}^{initial}_i$, and its new state, $\mathcal{X}^{new}_i$, while executing the DL re-training task. Since the system metrics can vary during the execution of the re-training task, we use the 95th percentile value statistic. The second step involves learning the performance degradation, i.e., the pressure on a background task, as a function of the new node state, defined by \texttt{EstExecTime} as shown in Equation~(\ref{eq:model4}).  
\begin{align}
&\mathcal{X}^{new}_{i} = \texttt{EstState}(\mathcal{X}^{initial}_{i}) \label{eq:model3}\\
& pressure^{a}_{i} = \texttt{EstExecTime}(\mathcal{X}^{new}_{i}) \label{eq:model4}
\end{align}

The models mentioned above are learned by first performing a sensitivity analysis to understand the importance/influence of the prospective features. Based on the candidate feature set obtained after the sensitivity analysis, regression models are learned. In \textit{Deep-Edge}, we use H2O’s AutoML framework~\cite{HomeML} to find the best hyper-parameter tuned algorithm.

\begin{figure*}
    \centering
    \includegraphics[width = 0.9\linewidth]{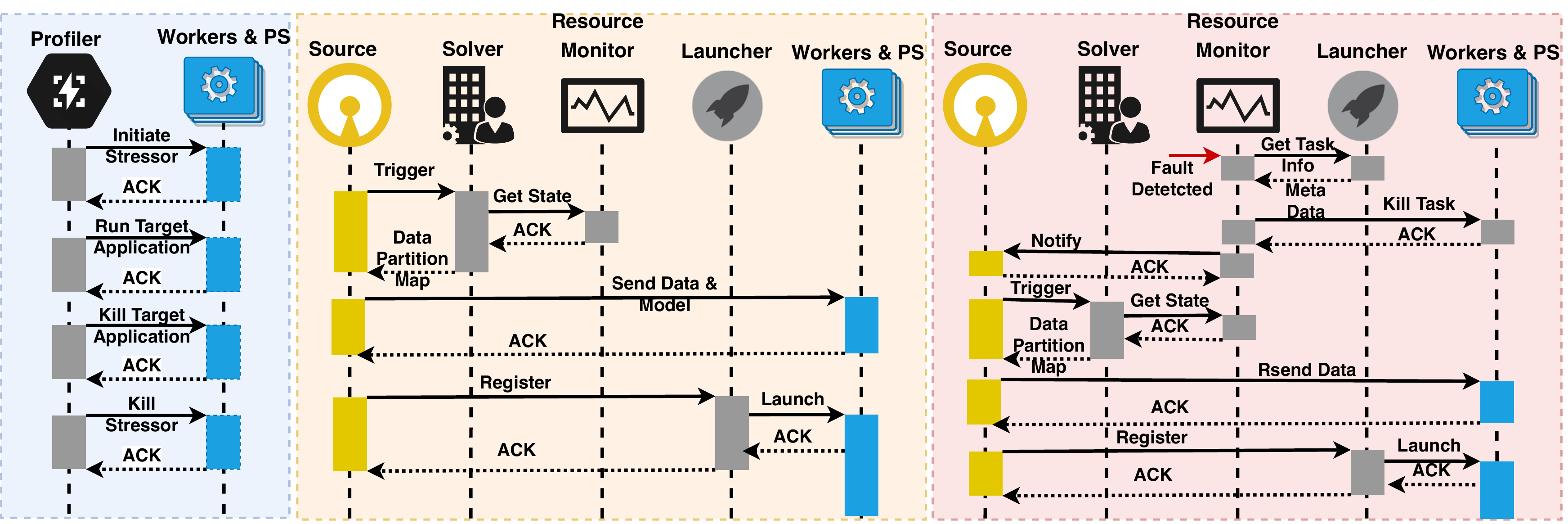}
    \caption{Event sequence diagram: Profiling (\textbf{Left}),  Job scheduling (\textbf{Middle}), Failure handling (\textbf{Right})}
    \label{fig:sequence}
\end{figure*}

\subsubsection{\textbf{Resource Scheduling}}
\label{subsec:scheduling}
Figure~\ref{fig:sequence} (Middle) shows the sequence of events pertaining to a new request of model update. A data store makes a request to DEM's solver endpoint to trigger the mode update. Upon the receipt of the request, the Solver gets the updated states of the workers, i.e., the number of prospective workers and their respective states (system metrics). Then, the Solver calculates the candidate worker nodes, data shards, and batch distribution using the scheduling strategy illustrated in Algorithm~\ref{alg:shardinglogic1}. The data and batch distributions are sent back to the data store to initiate data and base model transfer. After successful transmission, the data store registers the task information, such as the number of workers, data source, worker hostnames, data shards, and batch distribution, with the Launcher. In the end, the Launcher sends the acknowledgment back to the data store after successfully starting the DL re-training task on the selected workers. 

%\andy{Should we even use the term suboptimal because it may come across as a negative sounding term. Why can't we just say efficient because we already start the sentence with the word heuristic which implies that it is not an optimal algo. HS: done}
The heuristic presented in Algorithm~\ref{alg:shardinglogic1} provides an efficient solution to the optimization problem described in Section~\ref{sec:problem}. The input of the algorithm includes the workers' state map $\mathcal{X} = [\mathcal{X}_1, \mathcal{X}_2, \dots, \mathcal{X}_N]$, where  $\mathcal{X}_{i}$ = [$\mathcal{X}_i^{cpu}$, $\mathcal{X}_i^{gpu}$, $\mathcal{X}_i^{mem}$] represents the state of worker $w_i \in \mathcal{W}$, the parameter server's state $\mathcal{X}_{ps}$ = [$\mathcal{X}_{ps}^{cpu}$, $\mathcal{X}_{ps}^{gpu}$, $\mathcal{X}_{ps}^{mem}$], the number of data samples $\mathcal{M}$, the stopping threshold $\varepsilon$, and the maximum number of iterations $\tau$ attempted by the algorithm. The output of the algorithm is the data distribution $\widetilde{\mathcal{D}}$ and the batch size distribution $\widetilde{\mathcal{B}}$, such that $d_i \in \widetilde{\mathcal{D}}$ and $b_i \in \widetilde{\mathcal{B}}$ indicate, respectively, the data shard and batch size associated with worker $w_i \in \mathcal{W}$. Note that if any worker $w_k \in \mathcal{W}$ is not selected for the model update task, the algorithm will return $d_k= 0$ and $b_k = 0$ for that worker.

The heuristic computes the data shards and batch distribution in an iterative fashion, where the initial estimate of the size of all data shards is set to be $\infty$ [Lines 2-3]. Using the initial data shards and the memory utilization of a node, the corresponding batch size $b_i$ is calculated [Line 6] using the function \texttt{GetMaxBatchSize}, which provides the maximum batch size given the current memory utilization of the node while enforcing adherence of the memory constraint described in the optimization problem. With the batch size estimates, $t^{compute}_i$, $t^{update}_i$ and $t^{total}_i$ are calculated for all the workers [Lines 7-9]. A refined data shard $d_i$ for worker $w_i$ is then calculated based on $t^{total}_i$ to balance the workloads of all the workers [Line 10]. After calculating the refined data shards, all workers check for adherence for resource interference constraints [Lines 11-22]. A worker is removed from the DL cluster [Lines 19-21] if any of its background tasks will experience deadline violations [Lines 14-17]. The cycle [Lines 5-28] repeats until the data shards in two consecutive iterations are almost the same, i.e., the $\mathcal{L}^2$ norm is less than a threshold $\varepsilon$, or the maximum number of iterations $\tau$ have been reached [Lines 24-26].  

Then, we calculate the epoch time based on the data shards, which is given by the maximum time taken by any worker to process the data samples assigned [Line 29]. Note that, based on our proportional data sharding scheme, the difference between the epoch times from the different workers should be minimal (only due to rounding-off errors). If the resulting epoch time is better than the best one we have found so far, we remember the configuration as a potential solution [Lines 30-32]. Finally, to explore if better solutions are possible, we calculate the effect of removing the slowest worker (in terms of the total per-sample training time) on the overall epoch time [Lines 33-34]. As fewer workers are now present, this may affect the update time for each remaining worker, which will, in turn, affect the data shards and the epoch time. This process is performed iteratively until removing a worker no longer improves the overall epoch time, in which case the algorithm will eventually terminate [Line 36]. The final data shard size will then be given by the best one found so far.

\begin{algorithm}
    \scriptsize
    \caption{Scheduling Heuristic}\label{alg:shardinglogic1}
    \SetAlgoLined
    
    \textbf{Initialize}: $min\_t^* \leftarrow \infty$, $\widetilde{\mathcal{W}}\leftarrow \phi$, $\widetilde{\mathcal{B}}\leftarrow \phi$, $\widetilde{\mathcal{D}}\leftarrow \phi$; $Iter$ $\leftarrow$ 0\\
    $d_i = \infty,~\forall w_i \in \mathcal{W}$; \\
    $\mathcal{D} \leftarrow \{d_1, d_2, \dots, d_N\},~Iter \leftarrow 0$; \\
    \While{True}{
        \While{True}{
            $b_i \leftarrow \min(\textnormal{GetMaxBatchSize}(\mathcal{X}_i^{mem}), d_i),~\forall w_i \in \mathcal{W}$; \\
            $t_i^{compute} \leftarrow \textnormal{EstComputeTime}(\mathcal{X}_i, b_i),~\forall w_i \in \mathcal{W}$; \\
            $t^{update}_i \leftarrow \textnormal{EstUpdateTime}(\mathcal{X}_i, \mathcal{X}_{ps}, \mathcal{B}, |\mathcal{W}|), \forall w_i \in \mathcal{W}$;\\
            $t^{total}_i \leftarrow t^{compute}_i + t^{update}_i/b_i, \forall w_i \in \mathcal{W}$; \\
            $d_i = \frac{|\mathcal{M}|}{t^{total}_i \cdot \sum_{w_i\in \mathcal{W}} \frac{1}{t^{total}_i}}, \forall w_i \in \mathcal{W}$; \\
            \For {\textnormal{each} $w_i\in W$}{
                $flag \leftarrow$ False; \\
                \For{\textnormal{each} a $\in$ backApp$_i$ }{
                    \If{pressure$_{i}^{a}$ > $\delta_{i}^{a}$}{
                        $flag \leftarrow$ True;\\
                        break; \Comment Constraint violated\\
                    }
                }
                \If{$flag$}{
                    $\mathcal{W} \leftarrow \mathcal{W}\setminus w_i, b_i \leftarrow 0, d_i \leftarrow 0$; \Comment Remove worker\\
                }
            }
            $\mathcal{D}_{new} \leftarrow \{d_1, d_2, \dots, d_N\},~Iter$++; \\
            \If {($||\mathcal{D}_{new}$ - $\mathcal{D} ||_2\le \epsilon$) $\lor$ ($Iter == \tau$)}{
                break;
            }
            $\mathcal{D} \leftarrow \mathcal{D}_{new}$; \\
            
        }
        $epochTime = \max_{w_i\in \mathcal{W}} (d_i \cdot t^{total}_i)$; \\
        \uIf{$epochTime< min\_t^*$}{
            $min\_t^* \leftarrow epochTime$;\\
            $\widetilde{\mathcal{W}}\leftarrow \mathcal{W}$ , $\widetilde{\mathcal{B}}\leftarrow \mathcal{B}$ , $\widetilde{\mathcal{D}}\leftarrow \mathcal{D}$;\\
            $w_k \leftarrow \argmax_{w_i\in \mathcal{W}} (t^{total}_i)$; \Comment Find slowest worker \\
            $\mathcal{W} \leftarrow \mathcal{W} \setminus w_k, b_k \leftarrow 0, d_k \leftarrow 0$; \Comment Remove worker \\
            %$b_k \leftarrow 0$; \\
            %$d_k \leftarrow 0$;\\
        }
        \Else{break;}
    }
\end{algorithm}

\subsubsection{\textbf{Fault Tolerance}}
When a worker node experiences a failure while executing a model update task either due to process crash or node failure, the Resource Monitor in the DEM will detect such events and trigger the re-launching of the task. The DEM uses a \emph{three-strike rule}, i.e., a worker will not be considered part of the DL cluster if it has experienced at least three interruptions while running a model update task. Figure~\ref{fig:sequence} (Right) highlights the sequence of events as a result of worker failure. After detecting a worker failure, the Resource Monitor gets the DL task information such as data shards, batch distribution, worker hostnames, data source and task id from the Launcher, and kills the model update processes on the remaining workers. After the processes are killed, the Resource Monitor notifies the appropriate data source to re-trigger the model update task.

%------------------------------------------------------------------’

%--------------- SURVEY -------------------------------------
\section{Evaluation Results}
\label{sec:evaluation}
In this section, we present the evaluation results of different phases of the \textit{Deep-Edge} framework.  
%\vspace{-6pt}
\subsection{Experimental Setup}
\label{subsec:setup}
\textbf{Testbed}: Our testbed comprises one \emph{NVIDIA Jetson TX2} (256-core NVIDIA Pascal GPU, 8GB memory, Dual-Core NVIDIA Denver 2 64-Bit CPU and Quad-Core ARM Cortex-A57), three \emph{NVIDIA Jetson Nano} (128-core Maxwell GPU, Quad-Core ARM Cortex-A57, 4GB memory), and two \emph{Raspberry Pi 4} (Broadcom BCM2711, Quad-core Cortex-A72 (ARM v8), 4GB memory). These devices are connected by a layer 2 switch. One Raspberry Pi acts as a parameter server while the second acts as a data store and also hosts the DEM. 

\textbf{Workloads}: The model update task consists of updating a base DNN, namely Inception~\cite{szegedy2016rethinking}, with 3,855 data samples from the Caltech-256 dataset~\cite{griffin2007caltech} using MXNET~\cite{chen2015mxnet}. The base Inception model is created using transfer learning~\cite{tian2018continuum}, where the last layer of a pre-trained Inception model based on Imagenet~\cite{russakovsky2015imagenet} is replaced by a new layer. We trained the base model (last layer) with 2,700 data points to reach an accuracy of 60\%. The latency-critical background task is based on a distributed real-time computer vision application, which performs image reconstruction from multiple video streams where an initial image processing step is done in parallel on multiple edge devices. The image processing step involves identifying scale and rotation invariant descriptors (features) using Scale Invariant Feature Transform (SIFT)~\cite{lowe2004distinctive}. The latency-critical task constitutes executing SIFT on the acquired frame and sending the serialized SIFT features along with the original frame to an image stitching server over a UDP socket every 200ms. The size and resolution of the acquired frame are 56KB and 640$\times$320, respectively.       

\begin{figure*}
    \centering
    \includegraphics[width=0.9\textwidth]{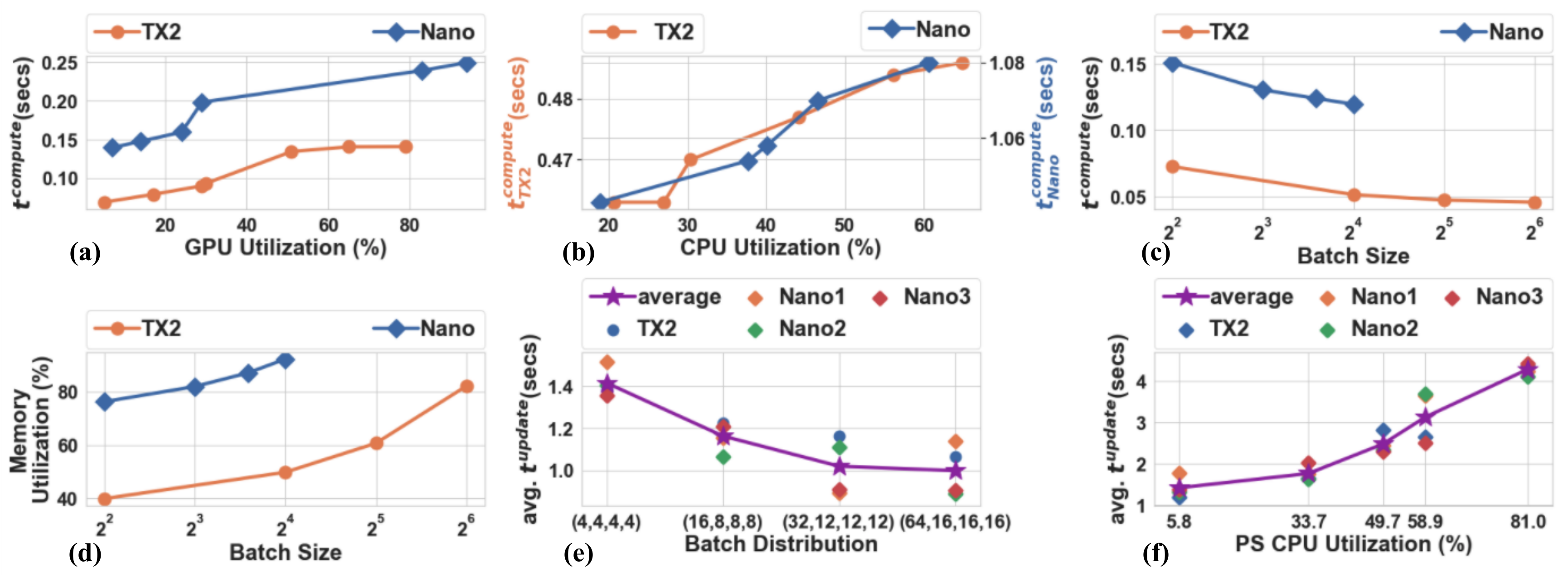}
    \caption{Sensitivity analysis}
    \label{fig:sensitivity}
\end{figure*}

\subsection{Performance Modeling}
\label{subsec:performanceresults}
We performed sensitivity analysis along the dimensions of GPU, CPU, and Memory utilization, as well as the number of workers and batch size. The following behaviors are observed:

\begin{itemize}
%\Romannum{1}) 
\item Increasing the GPU utilization increases the compute time $t^{compute}$ (Figure~\ref{fig:sensitivity}(a)).
\item Increasing the CPU utilization increases the compute time $t^{compute}$ (Figure~\ref{fig:sensitivity}(b)).
\item Increasing the batch size of the DL model update process decreases the compute time $t^{compute}$ (Figure~\ref{fig:sensitivity}(c)).
\item Increasing the batch size of the DL model update process increases the memory utilization (Figure~\ref{fig:sensitivity}(d)).
\item Increasing the batch size of the DL model update process decreases the update time $t^{update}$ (Figure~\ref{fig:sensitivity}(e)).
\item Increasing the CPU utilization of the parameter server increases the update time $t^{update}$ (Figure~\ref{fig:sensitivity}(f)). 
\item Increasing the number of workers increases the DL model update time $t^{update}$.
\item Increasing Memory utilization does not affect the throughput of the model update task. However, insufficient free memory can result in terminating a process by the OS.
\end{itemize}

Based on the above observations, we considered CPU utilization, GPU utilization, and batch size as the candidate features to learn the function \texttt{EstComputeTime} and added server CPU utilization, the number of workers along with batch distribution of all nodes to the above list as the set of features to learn the function \texttt{EstUpdateTime}. The function, \texttt{EstState}, has multiple outputs in nature, i.e., GPU, CPU, Memory, and a separate regressor is learned for each one of them. We extend the label with the feature name to indicate the individual regression model (for instance, \texttt{EstState}$^{mem}$ represents the memory regressor). \texttt{GetMaxBatchSize} uses \texttt{EstState}$^{mem}$ recursively to identify the maximum feasible batch size to run the model update task on a node. Finally, \texttt{EstExecTime} uses all system metrics of the node as features to predict the time to finish the latency-critical task.  We used H2O's AutoML framework~\cite{HomeML} to select the best regression algorithm as well as to perform hyperparameter tuning. Table~\ref{tab:results} highlights the number of data points, regression algorithm along with the accuracy for all learned functions. 
%As evident from the table, 
In particular, the gradient boosting based ensemble methods outperformed other algorithms with an average MAPE (mean absolute percentage error) of 3.433\% overall for all the learned functions.

\begin{table}
\centering
\scriptsize{

\begin{tabular}{c c c c c c}
\hline
\textbf{Device} & \textbf{Function}  & \textbf{Data-points}    & \textbf{Accuracy}   \\
                &                    & \textbf{(train/test)} &    \textbf{(MAPE)} \\
\hline
Nano & \texttt{EstComputeTime}   & 1386/264    & 1.668 $\pm$ 2.074 \\ 
Nano & \texttt{EstUpdateTime}    & 1386/264    & 9.617  $\pm$ 7.383 \\
Nano & \texttt{EstState}$^{gpu}$   & 1386/264    & 1.498  $\pm$ 3.9 \\ 
Nano & \texttt{EstState}$^{cpu}$    & 1386/264    & 5.21 $\pm$ 4.53 \\
Nano & \texttt{EstState}$^{mem}$    & 1386/264    & 1.508 $\pm$ 1.306 \\
Nano & \texttt{EstExecTime}    & 1386/264    & 1.73 $\pm$ 0.21 \\

\hline
TX2 & \texttt{EstComputeTime}   & 378/72    & 5.916 $\pm$ 6.721 \\ 
TX2 & \texttt{EstUpdateTime}    & 378/72    & 7.721  $\pm$ 6.598 \\
TX2 & \texttt{EstState}$^{gpu}$   & 378/72    & 0.509 $\pm$ 0.479 \\ 
TX2 & \texttt{EstState}$^{cpu}$    & 378/72    & 6.294 $\pm$ 5.468 \\
TX2 & \texttt{EstState}$^{mem}$    & 378/72    & 0.894 $\pm$ 0.704 \\
TX2 & \texttt{EstExecTime}    & 378/72    & 1.48 $\pm$ 0.12 \\

\hline
Pi & \texttt{EstState}$^{cpu}$    & 630/120    & 2.32 $\pm$ 1.58 \\
Pi & \texttt{EstState}$^{mem}$    & 630/120    & 1.70 $\pm$ 1.15 \\
\hline
\end{tabular}
}
\caption{Estimator results}
\label{tab:results}
\end{table}

\subsection{Scheduling}
\label{subsec:schedulingresults}
\begin{figure}
    \centering
    \includegraphics[width=\columnwidth]{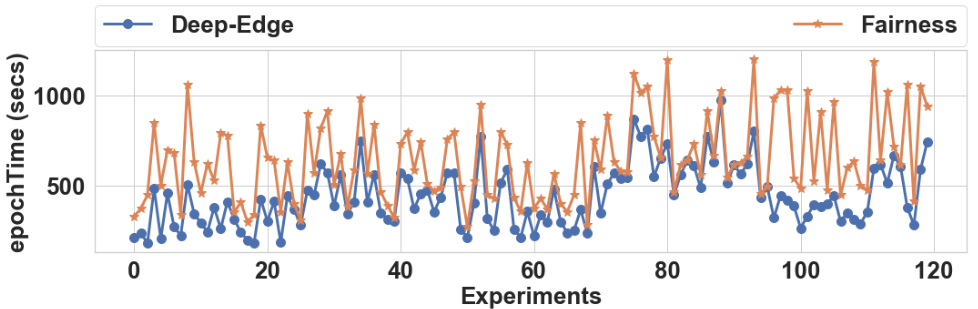}
    \caption{Epoch time distribution}
    \label{fig:results}
\end{figure}

\begin{figure}
    \centering
    \includegraphics[width=\columnwidth]{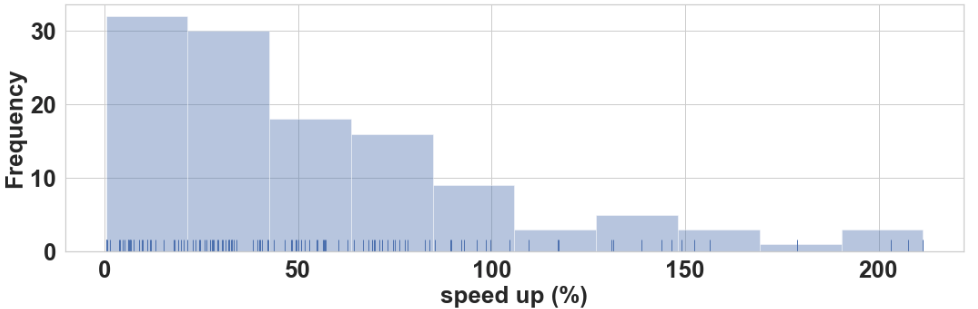}
    \caption{Speedup distribution}
    \label{fig:histogram}
\end{figure}

We compare the scheduling policy of \textit{Deep-Edge} with the fairness-based scheduler adopted in many resource managers, such as Hadoop~\cite{taylor2010overview}, Yarn~\cite{vavilapalli2013apache} and Mesos~\cite{hindman2011mesos}. We performed 120 random experiments, and in each experiment, all worker nodes are running the SIFT feature detector task along with some randomly selected stressors, such that the initial state of every node does not violate the deadline of the background application. Figure~\ref{fig:results} highlights the average epoch times observed when using the \textit{Deep-Edge} and the fairness schedulers. On average, the \textit{Deep-Edge} scheduler reduced the epoch time by 1.54 times compared to the fairness scheduler without violating any deadline. Figure~\ref{fig:histogram} shows a histogram of speedups for the DL task achieved by the \textit{Deep-Edge} scheduler over the fairness scheduler. We can see that the speedup can be as high as 200\%, and more than half of the experiments have a speedup of more than 50\%.    

% \subsection{Failure Handling}
% \label{subsec:failureresults}

% \subsection{Convergence Forecasting}
% \label{subsec:convergenceresults}

%------------------------------------------------------------------’

%--------------- Overview -------------------------------------
\section{Related Work }
\label{sec:survey}
This section provides a literature survey along the dimensions of interference aware resource management in edge devices and DL task scheduling.

\subsection{Performance \& Interference-aware Resource Management}
Resource interference is studied extensively in the literature, and different approaches are proposed to understand and quantify interference. Paragon~\cite{Delimitrou2013Paragon} presents an interference-aware job scheduler, in which an application’s performance is predicted using collaborative filtering. The performance prediction model is built by subjecting the target application by varying each resource stressor at a time. Authors in~\cite{zhao2013empirical} studied the impact of co-located application performance for a single multi-core machine and developed a piece-wise regression model using cache contention and bandwidth consumption of co-located applications as input features. The ESP project~\cite{mishra2017esp} also uses a regression model to predict performance interference for every possible co-location combination. Pythia~\cite{xu2018pythia} proposed a linear regression model approach for predicting combined resource contention by training on a small fraction of the large configuration space of all possible co-locations. Both ESP and Pythia assume that they have a priori information of all possible running workloads, based on which an interference model is created for a new application. PARTIES~\cite{chen2019parties} proposes a feedback-based controller to dynamically adjust resources between co-scheduled latency-critical applications using fine-grained monitoring and resource partitioning to guarantee the Quality-of-Service (QoS). INDICES~\cite{shekhar2017indices} proposes interference aware fog server selection using a gradient boosting based performance model of latency-critical applications. The authors extend the same approach in~\cite{shckhar2019urmila} to offload a latency-critical task between fog and edge devices while considering user mobility. However, none of these approaches are designed for distributed tasks such as DL model re-training and do not consider GPU utilization.

\subsection{Deep Learning Task Scheduling}
There are several approaches in the scientific literature for resource allocation to achieve a variety of objectives in cloud settings such as Borg~\cite{verma2015large}, Coral~\cite{jalaparti2015network} and TetriSched~\cite{tumanov2016tetrisched}, Morpheus~\cite{jyothi2016morpheus}. However, the schedulers mentioned above are not designed for DL workloads. There are recent research efforts on GPU sharing for machine learning tasks. Baymax~\cite{chen2016baymax} explores GPU sharing as a way to mitigate both queuing delay and resource contention. Following that, Prophet~\cite{chen2017prophet} proposes an analytical model to predict the performance of GPU workloads. Gandiva~\cite{xiao2018gandiva} aims GPU time-sharing in shared GPU clusters through check-pointing at low GPU memory usage of the training job. CROSSBOW~\cite{koliousis2019crossbow} proposes a dynamic task scheduler to automatically tune the number of workers to speed up the training and to use the infrastructure optimally. Optimus~\cite{peng2018optimus} also dynamically adjusts the number of workers and parameter servers to minimize the training completion time while achieving the best resource efficiency. SLAQ~\cite{zhang2017slaq} targets the training quality of experimental ML models instead of models in production. It adopts an online fitting technique similar to Optimus to estimate the training loss of convex algorithms. Dorm~\cite{sun2017towards} uses a utilization-fairness optimizer to schedule jobs. However, these approaches are not applicable for edge clusters as none of them considers resource interference while allocating heterogeneous resources for the DL model update task.

%--------------- CONCLUSION -------------------------------------
\section{Conclusion and Future Work}
\label{sec:conclusion}
This paper presents \emph{Deep-Edge}, an interference aware DL model update framework for the edge devices that minimizes the re-training time by intelligently distributing data among edge nodes while adhering to the latency constraints of the background applications.  We described different components of the framework and showed its efficacy by validating it against a realistic case study. 

In the future, we would like to extend this work in three dimensions: 1) improving performance and interference models by adding more features such as memory and disk bandwidth; 2) adding support for Multi-Process Service (MPS) based GPU workloads; 3) including parameter server load balancing as part of the scheduling problem; and 4) considering a more diverse set of edge devices such as TPUs and FPGAs.

%--------------- ACKNOWLEDGMENT -------------------------------------

% use section* for acknowledgment

\section*{Acknowledgment}
\footnotesize
This work was supported in part by AFOSR DDDAS FA9550-18-1-0126 program, Siemens Corporate Technology and NIST. Any opinions, findings, and conclusions or recommendations expressed in this material are those of the author(s) and do not necessarily reflect the views of these sponsors.

\bibliography{references}
\bibliographystyle{IEEEtran}

% that's all folks
\end{document}